\documentclass[twocolumn,prb]{revtex4-1}
\usepackage[T1]{fontenc}
\usepackage[latin9]{inputenc}
\usepackage{pmboxdraw}
\usepackage{amsmath}
\usepackage{graphicx}

\makeatletter
\usepackage{lipsum}
\usepackage{color}
\usepackage{hyperref}
\hypersetup{
     colorlinks   = true,
     citecolor    = blue
}

\makeatother

\begin{document}

\title{\noindent Edge on-site potential effects in a honeycomb topological
magnon insulator}

\author{\noindent Pierre A. Pantale{\'o}n}

\affiliation{\noindent Theoretical Physics Division, School of Physics and Astronomy,
University of Manchester, Manchester M13 9PL, United Kingdom}

\author{\noindent Y. Xian}

\affiliation{\noindent Theoretical Physics Division, School of Physics and Astronomy,
University of Manchester, Manchester M13 9PL, United Kingdom}
\begin{abstract}
The difference between the edge on-site potential and the bulk values
in a magnonic topological honeycomb lattice leads to the formation
of edge states in a bearded boundary, and the same difference is found to be 
the responsible for the absence of edge states in a zig-zag termination.
In a finite lattice, the intrinsic on-site interactions along the
boundary sites generate an effective defect and Tamm-like edge states
appear for both zig-zag and bearded terminations. If a non-trivial
gap is induced, Tamm-like and topologically protected edge states
appear in the band structure. The effective defect can be strengthened
by an external on-site potential and the dispersion relation, velocity
and magnon-density of the edge states become tunable. 
\end{abstract}
\maketitle

\section{Introduction }

Many important phenomena in condensed matter physics are related to
the formation of edge or surface states along the boundary of finite-sized
materials. Their existence has been commonly explained as the manifestation
of Tamm \cite{Tamm} or Shockley \cite{Shockley1939} mechanisms.
In recent years it has been revealed that the edge states in the so-called
topological insulators\cite{Hasan2010} are related to the bulk properties
\cite{Hatsugai1993,Hatsugai1993a}. One such property is characterized
by an insulating bulk gap and conducting gapless topologically protected
edge states that are robust against internal and external perturbations
\cite{Delplace2011,Malki2017}. 

Edge states in topological magnon insulators have also attracted a
lot of attention recently \cite{Matsumoto2011,Zhang2013,Mook2014,Owerre2016d}.
The magnons are the quantized version of spin-waves \cite{Bloch1930,Holstein1940},
which are collective propagation of precessional motion of the magnetic
moments in magnets. The intrinsic bosonic nature of the magnons allow
them to propagate over long distances without dissipation by Joule
heating \cite{Buttner2000,Kajiwara2010a}. Similar to spintronics\cite{Pulizzi2012},
the study of the edge magnons will enrich the potential of magnonics,
exploiting spin-waves for information processing\cite{Kruglyak2006,Kruglyak2010,Chumak2015,Chumak2017}.
For this purpose the basic understanding of the magnon behavior in
different lattice structures and the precise control of their properties
are urgently called for. 

The magnon hall effect was observed in the ferromagnetic insulator
$\textrm{Lu}_{2}V_{2}O_{7}$ \cite{Onose2010}, in the K{\'a}gome
ferromagnetic lattice \cite{Chisnell2015}, in $\mathrm{Y}_{3}\mathrm{Fe}{}_{5}\mathrm{O}_{3}$
(YIG) ferromagnetic crystals \cite{Madon2014,Tanabe2016}, and have
also been studied in the Lieb \cite{Cao2015} and the honeycomb ferromagnetic
lattices \cite{Owerre2016d}. Interestingly, it has been shown that
a ferromagnetic Heisenberg model with a Dzialozinskii-Moriya interaction
(DMI) on the honeycomb lattice realizes magnon edge states similar
to the Haldane model for spinless fermions \cite{Owerre2016d} and
the Kane-Mele model for electrons \cite{Kim2016a}. By a topological
approach, it has been shown that a non-zero DMI makes the band structure
topologically non-trivial and by the winding number of the bulk Hamiltonian,
gapless edge states which cross the gap connecting the regions near
the Dirac points has been predicted \cite{Owerre2016d}. The thermal
Hall effect \cite{Owerre2016c} and spin Nernst effect \cite{Kim2016a}
have also been predicted for this magnetic system. By a direct tight
binding formulation in an strip geometry, it was shown that the edge
states in a lattice with a zigzag termination closely resembles their
fermionic counterpart only if an external on-site potential is introduced
at the outermost sites \cite{Pantaleon2017}. Furthermore, the lattice
with armchair termination has additional edge states to those predicted
by a topological approach. Such edge states were found to be strongly
dependent to edge on-site potentials \cite{Pantaleon2017a}. On the
other hand, in a semi-infinite ferromagnetic square lattice, a renormalization
of the on-site contribution along the boundary gives rise to spin-wave
surface states \cite{DeWames1969,Hoffmann1973,Puszkarski1973} and
most recent experiments in photonic lattices have observed unconventional
edge states in a honeycomb lattice with bearded \cite{Plotnik2014},
zigzag and armchair \cite{Mili2017} boundaries, which are not present
in the fermionic graphene. In addition, Tamm-like edge states were
also observed in a K{\'a}gome acoustic lattice \cite{Ni2017}. These
unconventional edge states are found to be related to the bosonic
nature of the quasi-particles in the lattice whose model hamiltonians
contains on-site interaction terms. 

In this work, we explore in some detail the magnon edge states in
a honeycomb lattice with a DMI and an external on-site potential along
the outermost sites. Extending the results of our previous work\cite{Pantaleon2017,Pantaleon2017a},
where we found that the edge states depends strongly on the external
on-site potential, here, we present the general approach applied to
both zig-zag and bearded boundaries. We also derive analytical expressions
for both energy spectrum and wavefunctions, where the dependence in
the on-site potential appears explicitly. In a lattice with a boundary,
the interaction terms along the outermost sites differ from the bulk
values. In agreement with recent experiments\cite{Plotnik2014}, such
difference plays the role of an effective defect and gives rise to
Tamm-like edge states \textSFx{} the type of edge states generated
by an strong perturbation due to an asymmetric termination of a periodical
potential\cite{Tamm}. Therefore, with regard to the fact that the
on-site interaction terms along the boundary play an important role
in a bosonic lattice with a boundary then additional edge states are
found in the band structure.The intrinsic on-site potential at the
outermost sites generate novel edge states for both terminations.
We found that the effective defect can be strengthened by an external
on-site potential and this can be used to tune up the dispersion and
velocity of the edge states present in the system. For the both considered
boundaries, we present a simple diagram with which the number of magnon
edge states can be predicted. In addition, if a non-trivial gap is
induced, the edge state band structure is found to be strongly dependent
to the on-site interactions. The tight binding formulation which we
have implemented in this work facilitates extraction of analytical
solutions of both energy spectrum and wavefunctions for better physical
understanding. All our results are in agreement with direct numerical
calculations.

\section{tight-binding model on the honeycomb lattice \label{sec: Tight Binding Model}}

In this section, we briefly present the general approach for the study
of the edge states with an arbitrary external on-site potential and
with a DMI. 

\subsection{Harper's equation\label{subsec: Harper's equation}}

\begin{figure}
\begin{centering}
\includegraphics[scale=0.53]{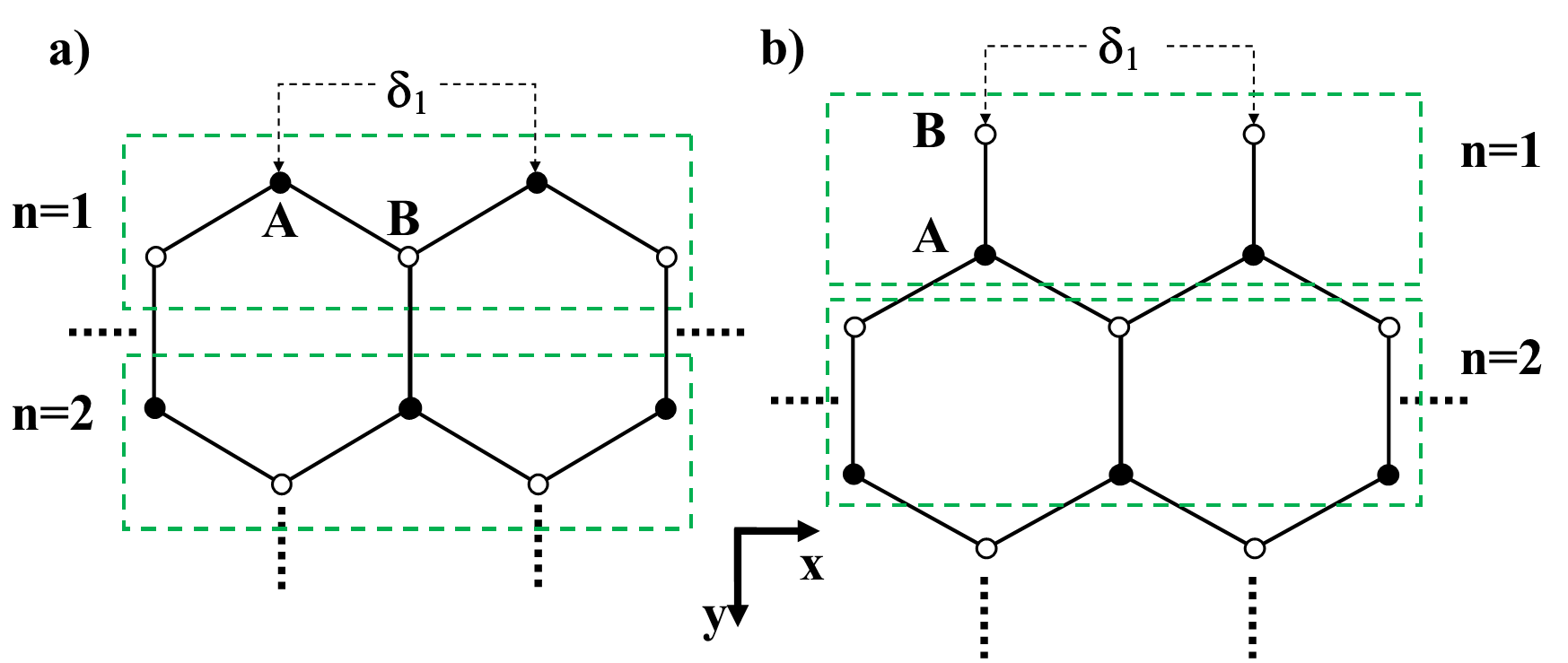}
\par\end{centering}
\caption{{\small{}(Color online) Squematics of the a) zig-zag and b) bearded
boundaries on the honeycomb lattice. The sub-lattices are labeled
by $A$ and $B$. The external on-site potential $\delta_{1}$ is
applied at the outermost sites. Here, $n$ is a index row along the
$y$ direction perpendicular to the boundary. \label{fig:Figure1}}}
\end{figure}

The bosonic tight-binding Hamiltonian on the honeycomb lattice, derived
from a linear spin-wave approximation to the Heisenberg model, is
given by
\begin{eqnarray}
H & = & -JS\sum_{\left\langle i,j\right\rangle }\left(a_{i}b_{j}^{\dagger}+a_{i}^{\dagger}b_{j}-a_{i}^{\dagger}a_{i}-b_{j}^{\dagger}b_{j}\right)+H_{D},\label{eq: Bosonic Hamiltonian}
\end{eqnarray}
where $a_{i}$ and $b_{j}$ are bosonic operators of the two sub-lattices,
$\left\langle i,j\right\rangle $ indicates a nearest-neighbor (NN)
coupling with isotropic ferromagnetic coupling constant $J\left(>0\right)$
and $S$ the quantum number from the original Heisenberg model. The
second term, $H_{D}=H_{D,A}+H_{D,B}$, is the DMI contribution, in
particular, for the $A$-sublattice is given by
\begin{equation}
H_{D,A}=iDS\sum_{\left\langle \left\langle i,j\right\rangle \right\rangle }\varrho_{i,j}\left(a_{i}a_{j}^{\dagger}-a_{i}^{\dagger}a_{j}\right),\label{eq: HDMI sub A}
\end{equation}
where $D$ is the DMI strength, $\left\langle \left\langle i,j\right\rangle \right\rangle $
runs over the next-nearest-neighbor (NNN) sites and the hopping term
$\varrho_{ij}=\pm1$ depending of the orientation of the two NNN sites
\cite{Kane2005}, $H_{D,B}$ is similar for the $B$-sublattice. The
Hamiltonian in Eq. (\ref{eq: Bosonic Hamiltonian}) is the bosonic
equivalent to the Haldane model \cite{Haldane1988}, where the NNN
complex hopping, Eq. (\ref{eq: HDMI sub A}), breaks the lattice inversion
symmetry and makes the band structure topologically non-trivial. To
analyze the edge states we consider a lattice with an open boundary
along the $x$ direction and semi-infinite in the $y$ direction,
Fig. (\ref{fig:Figure1}). In the linear spin-wave approximation,
by denoting wavefunctions on two sub-lattices of the honeycomb lattice
as $\psi_{A,n}$ and $\psi_{B,n}$, respectively, the Harper's equation
\cite{Harper1955} provided by the Hamiltonian in Eq. (\ref{eq: Bosonic Hamiltonian})
can be written as \textbf{
\begin{eqnarray}
3\psi_{A,n}-J_{1}\psi_{B,n}-J_{2}\psi_{B,n-1}+f_{A,n} & = & \varepsilon\psi_{A,n},\nonumber \\
-J_{1}\psi_{A.n}-J_{2}\psi_{A,n+1}+3\psi_{B,n}-f_{B,n} & = & \varepsilon\psi_{B,n},\label{eq:  Harper Equation General}
\end{eqnarray}
}where $n$ is a row index in the $y$ direction perpendicular to
the boundary. In the above equation, the DMI is given by $f_{l,n}=J_{3}\psi_{l,n}-J_{4}\left(\psi_{l,n+1}+\psi_{l,n-1}\right)$,
with $l$ $(=A,B)$ a sublattice index. Furthermore, if $k$ is the
momentum in the $x$ direction, the hopping amplitudes for the lattice
with a zig-zag boundary are given by: $J_{1}=2\cos\left(\sqrt{3}k/2\right)$,
$J_{2}=1$, $J_{3}=2\,D^{\prime}\,\sin\left(\sqrt{3}\,k\right)$,
$J_{4}=2\,D^{\prime}\,\sin\left(\sqrt{3}k/2\right)$ and $D^{\prime}=D/J$.
In addition, the simple replacements of $J_{1}\rightarrow J_{2}$
and $J_{2}\rightarrow J_{1}$ in the Eq. (\ref{eq:  Harper Equation General})
provide us the corresponding Harper's equation for the lattice with
a bearded boundary. 

\subsection{Effective Hamiltonian for the edge states \label{subsec:Effective Hamiltonian}}

The Harper's equation, Eq. (\ref{eq:  Harper Equation General}),
can be simplified if we assume a decaying Bloch wavefunction in the
$y$ direction of the form, $\psi_{l,n}=z^{n}\psi_{l}$, where $l$
labels each sublattice and the Bloch phase factor $z$ a complex number
\cite{Celic1996,Pavkov2001}. The effective Hamiltonian for the edge
state can be written with the decaying wavefunction as $H_{ef}\psi_{l,n}=\varepsilon\psi_{l,n}$,
where 
\begin{equation}
H_{ef}=\left[\begin{array}{cc}
3+J_{3}-J_{4}\Delta & -w\left(J_{1}+J_{2}z^{-1}\right)\\
-w^{-1}\left(J_{1}+J_{2}z\right) & 3-J_{3}+J_{4}\Delta
\end{array}\right],\label{eq: Effective Hamiltonian for the edge state}
\end{equation}
and $\Delta=z+z^{-1}$. In the above equation, the factor $w$ takes
into account the bearded $(w=z)$ and zig-zag $(w=1)$ boundaries.
The non-trivial solution for the eigenstates of $H_{ef}$ gives rise
to the secular equation 
\begin{equation}
J_{4}^{2}\Delta{}^{2}-\left(2J_{3}J_{4}-J_{1}J_{2}\right)\Delta-\varepsilon_{r}^{2}+J_{1}^{2}+J_{2}^{2}+J_{3}^{2}=0,\label{eq: Quadratic equation general}
\end{equation}
where $\varepsilon_{r}=\varepsilon-3$. Note that such polynomial
in $\Delta$ is the same for the both considered boundaries. For a
given momentum $k$ and energy $\varepsilon$, the solutions of Eq.
(\ref{eq: Quadratic equation general}) are the Bloch phase factors
$z_{\nu}$, $\nu=1,..,4$. In particular, for the infinite system,
the Fourier transform in the $y$ direction is the solution $z=e^{\pm\frac{3}{2}ik_{y}}$
which corresponds to Bloch extended states. In the case of a lattice
with a boundary, the solutions of Eq. (\ref{eq: Quadratic equation general})
satisfying $\left|z_{\nu}\right|=1$ determine the bulk band structure
{[}See. Fig. (\ref{fig: Figure2}){]}. The states with $\left|z_{\nu}\right|\neq1$
decay/grow exponentially in space, and they can be used to describe
the edge states with the appropriate boundary conditions. 

The factors $z_{\nu}$ and $z_{\nu}^{-1}$ in the Eq. (\ref{eq: Quadratic equation general})
always appear in pairs. Since we require a decaying (evanescent) wave
from the boundary, setting the condition $\left|z_{\nu}\right|<1$
implies that the general solution for the edge states can be written
as a linear combination of the form,
\begin{equation}
\psi_{l,n}=c_{1}z_{1}^{n}\psi_{l}^{\left(1\right)}+c_{2}z_{2}^{n}\psi_{l}^{\left(2\right)},\label{eq: General Linear Combination}
\end{equation}
where the coefficients $c_{i}$ are determined by the boundary conditions.
In the above equation, $\psi_{l}^{(\nu)}$, $\nu=1,2$, is an eigenvector
of $H_{ef}$ corresponding to the $\nu-th$ solution. To obtain the
edge state energy spectrum, the wavefunctions, Eq. (\ref{eq: General Linear Combination}),
must satisfy the boundary conditions. This will be described in the
following sections.

\section{boundary conditions and the edge states\label{sec: Boundary conditions and edge states}}

In this section, the boundary conditions for both zig-zag and bearded
boundaries are obtained. By the secular Eq. (\ref{eq: Effective Hamiltonian for the edge state})
and the boundary conditions, we derive the analytical expressions
for the edge state energy spectrum and wavefunctions for non-zero
DMI. Please refer to the appendix \ref{sec: Analytical D0} for the
solutions with zero DMI.

\subsection{Zig-zag boundary }

In our previous work\cite{Pantaleon2017}, we derive the equations
for the energy and the wavefunctions considering a fixed on-site potential
$\delta_{1}=1$, where the edge state energy spectrum and the wavefunctions
closely resembles the fermionic graphene. Here, we will just summarize
the derivation with the notation in this paper and extending the formalism
to arbitrary external on-site potentials. 

Due to the open zig-zag boundary, the on-site potential along the
boundary is different from that in the bulk. Then, the Harper's equation,
Eq. (\ref{eq:  Harper Equation General}), at $n=1$ must be modified.
Considering the missing bonds along the outermost $A$ site, the coupled
Harper's equation at $n=1$ is written as,\textbf{
\begin{align}
\left(2-\delta_{1}\right)\psi_{A,1}-J_{1}\psi_{B,1}+f_{A,1} & =\varepsilon\psi_{A,1},\nonumber \\
3\psi_{B,1}-\left(J_{1}\psi_{A.1}+J_{2}\psi_{A,2}\right)-f_{B,1} & =\varepsilon\psi_{B,1},\label{eq: Harper zigzag n1}
\end{align}
}where the external on-site potential $\delta_{1}$ is introduced
and $f_{l,1}=J_{3}\psi_{l,1}-J_{4}\psi_{l,2}$. In the the above equation,
the total energy at each sublattice is given by the on-site contribution
(first term), the NN contribution (second term) and the DMI (third
term). Contrasting the Eq. (\ref{eq:  Harper Equation General}) and
Eq. (\ref{eq: Harper zigzag n1}), we obtain the zig-zag boundary
conditions 
\begin{align}
\left(1-\delta_{1}\right)\psi_{A,1}-J_{4}\psi_{A,0} & =0,\label{eq: zig zag boundary condition}\\
\psi_{B,0} & =0,\nonumber 
\end{align}
for the edge state wavefunctions in Eq. (\ref{eq: General Linear Combination}).
Unlike the equivalent fermionic model\cite{Doh2013} where the wavefunctions
of both sub-lattices vanish at $n=0$. The Eq. (\ref{eq: zig zag boundary condition})
contains two additional terms; the intrinsic and the external on-site
potential. As we will shown in the following sections, such extra
terms have important effects. From the Eq. (\ref{eq: General Linear Combination})
and Eq. (\ref{eq: zig zag boundary condition}) the non-trivial solution
for the coefficients $c_{i}$ provide us the following self-consistent
equation for the edge states, \begin{widetext}

\begin{equation}
\varepsilon=3+J_{3}-J_{4}\left\{ \frac{\left[\left(\delta_{1}-1\right)J_{1}-J_{2}J_{4}\right]\left(z_{1}+z_{2}\right)+\left[J_{2}\left(\delta_{1}-1\right)+J_{1}J_{4}\right]\left(1-z_{1}z_{2}\right)}{\left(\delta_{1}-1\right)J_{1}z_{1}z_{2}-J_{2}J_{4}}\right\} ,\label{eq: Energy autoconsistent zigzag}
\end{equation}
\end{widetext} with $z_{1}$ and $z_{2}$ the two decaying solutions
of the Eq. (\ref{eq: Quadratic equation general}). The corresponding
edge state wavefunctions are given by 
\begin{eqnarray}
\psi_{A,n} & = & c_{1}\left(z_{1}^{n}-\alpha z_{2}^{n}\right)\psi_{A}^{\left(1\right)},\nonumber \\
\psi_{B,n} & = & c_{1}\left(z_{1}^{n}-z_{2}^{n}\right)\psi_{B}^{\left(1\right)},\label{eq: Wavefunctions zig-zag}
\end{eqnarray}
where $c_{1}$ is a normalization term, and 
\begin{equation}
\alpha=\frac{\left(1-\delta_{1}\right)z_{1}-J_{4}}{\left(1-\delta_{1}\right)z_{2}-J_{4}},\label{eq: alpha factor zig-zag}
\end{equation}
contains the contribution of the external on-site potential in the
wavefunction. For a given momentum $k$, external potential $\delta_{1}$
and non-zero DMI, the Eq. (\ref{eq: Energy autoconsistent zigzag})
is an implicit equation for the energy $\varepsilon$ and can be solved
numerically. The Eq. (\ref{eq: Quadratic equation general}), Eq.
(\ref{eq: Energy autoconsistent zigzag}) and Eq. (\ref{eq: Wavefunctions zig-zag})
provide us a full description for the edge states energy spectrum
and wavefunctions, which will be described in the Sec. \ref{sec: Energy Spectrum}. 

\subsection{Bearded boundary}

Similar to the zig-zag case, by modifying the Harper's equation at
$n=1$ to take into account the missing sites, the boundary conditions
for the wavefunctions in Eq. (\ref{eq: General Linear Combination})
are given by
\begin{align}
\left(2-\delta_{1}\right)\psi_{B,1}+J_{4}\psi_{B,0} & =0,\label{eq: General Bearded boundary condition}\\
\psi_{A,0} & =0.\nonumber 
\end{align}
From the Eq. (\ref{eq: General Linear Combination}) and Eq. (\ref{eq: General Bearded boundary condition})
the non-trivial solution for the coefficients $c_{i}$ can also be
obtained. However, by a closer inspection of the boundary conditions
and the Harper's equation, we found that the simple replacements:
$J_{1}\rightarrow J_{2}$, $J_{2}\rightarrow J_{1}$, $J_{3}\rightarrow-J_{3}$,
$J_{4}\rightarrow-J_{4}$ and $\delta_{1}\rightarrow\delta_{1}+1$
in the Eq. (\ref{eq: Energy autoconsistent zigzag}), provide us the
required self-consistent equation for the edge state energy spectrum.
The wavefunctions satisfying the boundary conditions, Eq. (\ref{eq: General Bearded boundary condition}),
are given by 
\begin{eqnarray}
\psi_{A,n} & = & c_{1}\left(z_{1}^{n}-z_{2}^{n}\right)\psi_{A}^{\left(1\right)},\nonumber \\
\psi_{B,n} & = & c_{1}\left(z_{1}^{n}-\alpha^{\prime}z_{2}^{n}\right)\psi_{B}^{\left(1\right)},\label{eq: General Wavefunctions Bearded}
\end{eqnarray}
where $c_{1}$ is a normalization term and 
\begin{equation}
\alpha^{\prime}=\frac{\left(2-\delta_{1}\right)z_{1}+J_{4}}{\left(2-\delta_{1}\right)z_{2}+J_{4}}.\label{eq: alpha factor bearded}
\end{equation}
The Eq. (\ref{eq: Quadratic equation general}) and the self-consistent
equation obtained by the Eq. (\ref{eq: General Bearded boundary condition})
and the Eq. (\ref{eq: General Wavefunctions Bearded}) provide us
a full description for the edge state energy spectrum and wavefunctions.
For an arbitrary external on-site potential and zero DMI, the $k-$dependence
of $\varepsilon(k)$ and the explicit solutions for the decaying $z$
factors are obtained in the appendix \ref{sec: Analytical D0}. 
\begin{figure}
\begin{centering}
\includegraphics[scale=0.187]{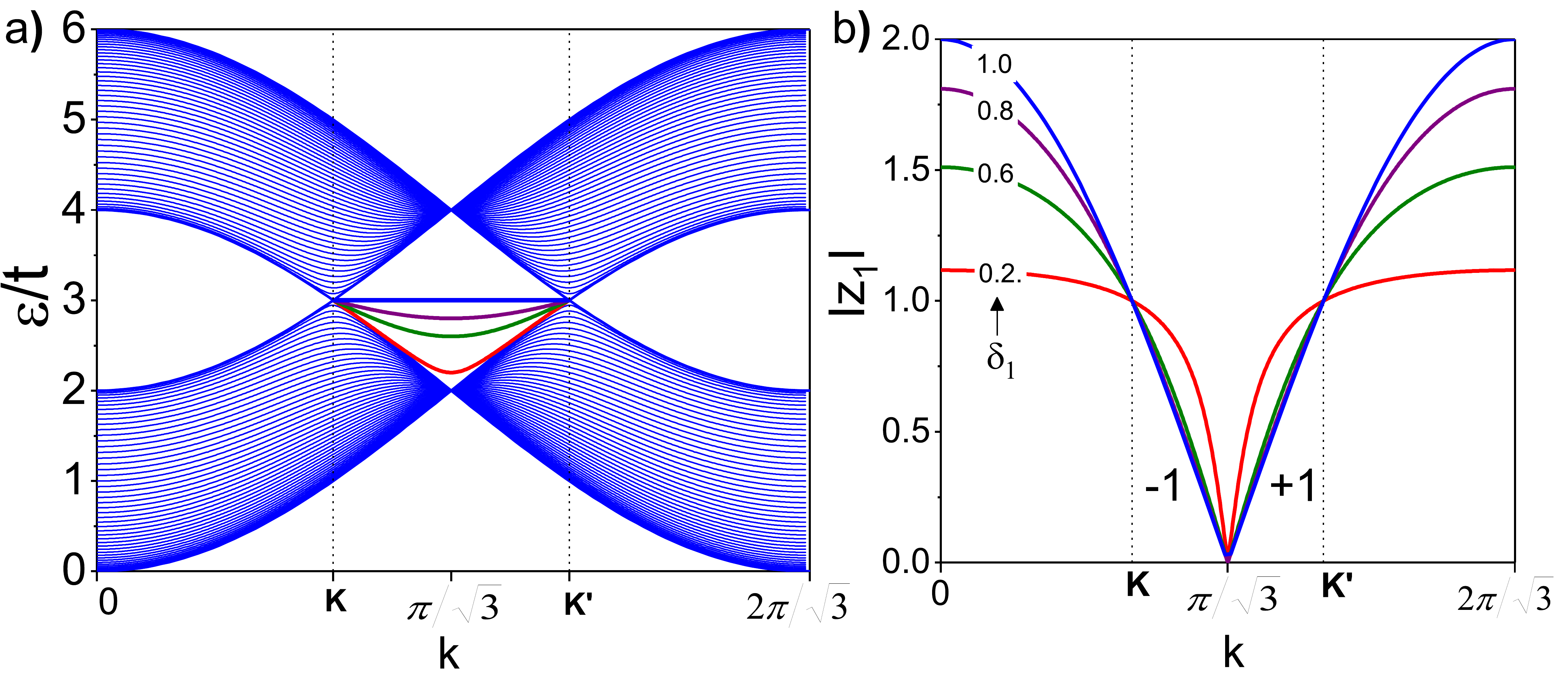}
\par\end{centering}
\caption{{\small{}(Color online) a) Edge state dispersion relations induced
by $\delta_{1}$ at the zig-zag open boundary are shown for $\delta_{1}=0.2,\,0.6,\,0.8$
and $1.0$ (from the curved to the flat band). For $\delta_{1}=0$
there are not edge states. The blue region is the bulk continuum where
all the factors $\left|z_{\upsilon}\right|=1$ in Eq. (\ref{eq: Quadratic equation general}).
b) Modulus of the decaying factors for the corresponding edge states,
here $\pm1$ is the sign of $z_{1}$. \label{fig: Figure2}}}
\end{figure}

\section{energy spectrum and wavefunctions\label{sec: Energy Spectrum}}

\subsection{Zero DMI}

In a fermionic lattice with a boundary, it is well known that there
are flat edge states connecting the two Dirac points, $\boldsymbol{K}\rightarrow\boldsymbol{K^{\prime}}$,
in a lattice with a zig-zag boundary\cite{Fujita1996}, where in a
lattice with a bearded boundary \cite{Klein1994}, the flat edge state
is connecting the complementary region, $\boldsymbol{K^{\prime}}\rightarrow\boldsymbol{K}$.
In the equivalent bosonic models, some differences are expected due
to the contribution of the on-site interactions along the boundary
sites. 

\begin{figure}
\begin{centering}
\includegraphics[scale=0.04]{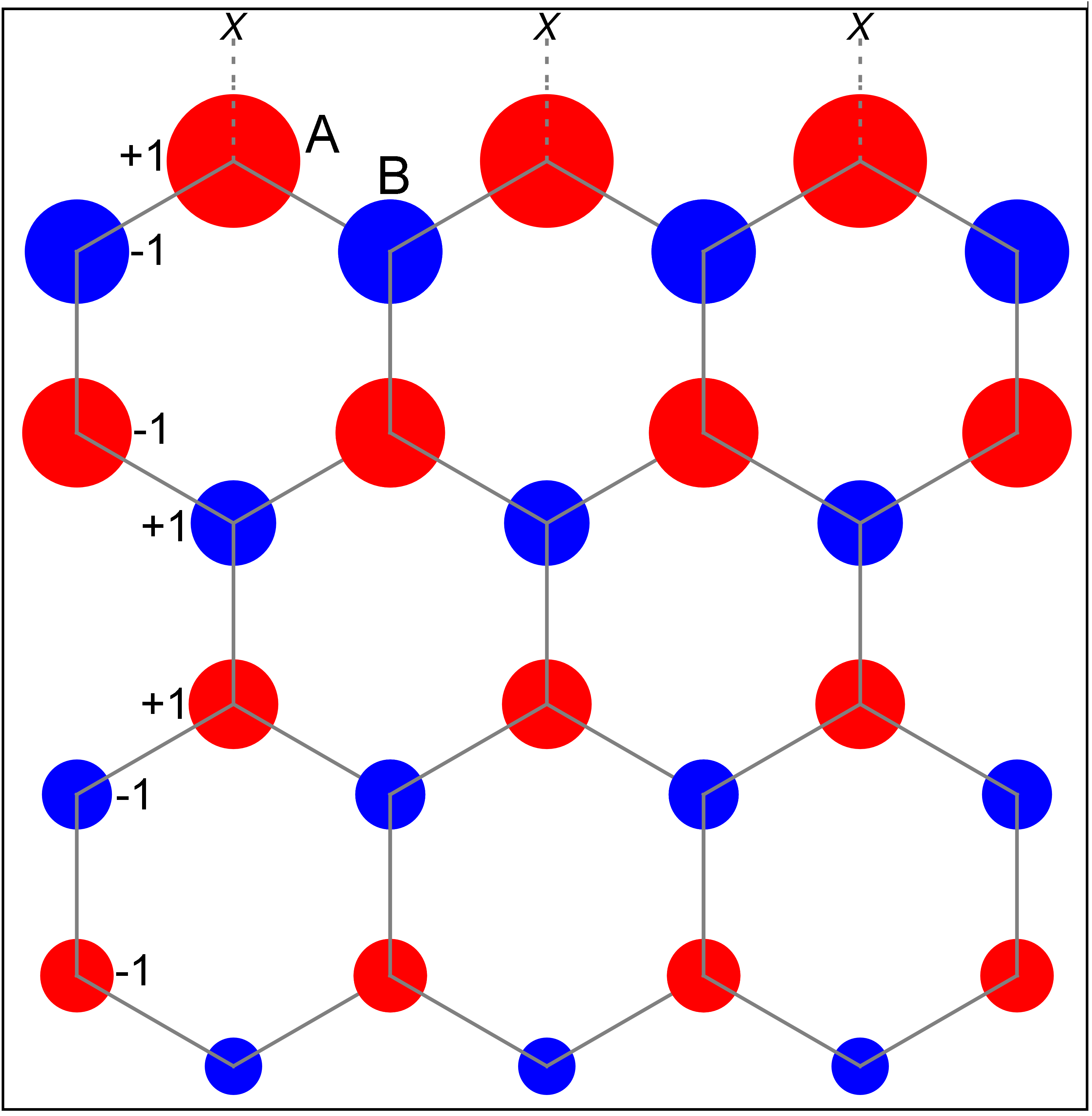}~\includegraphics[scale=0.0397]{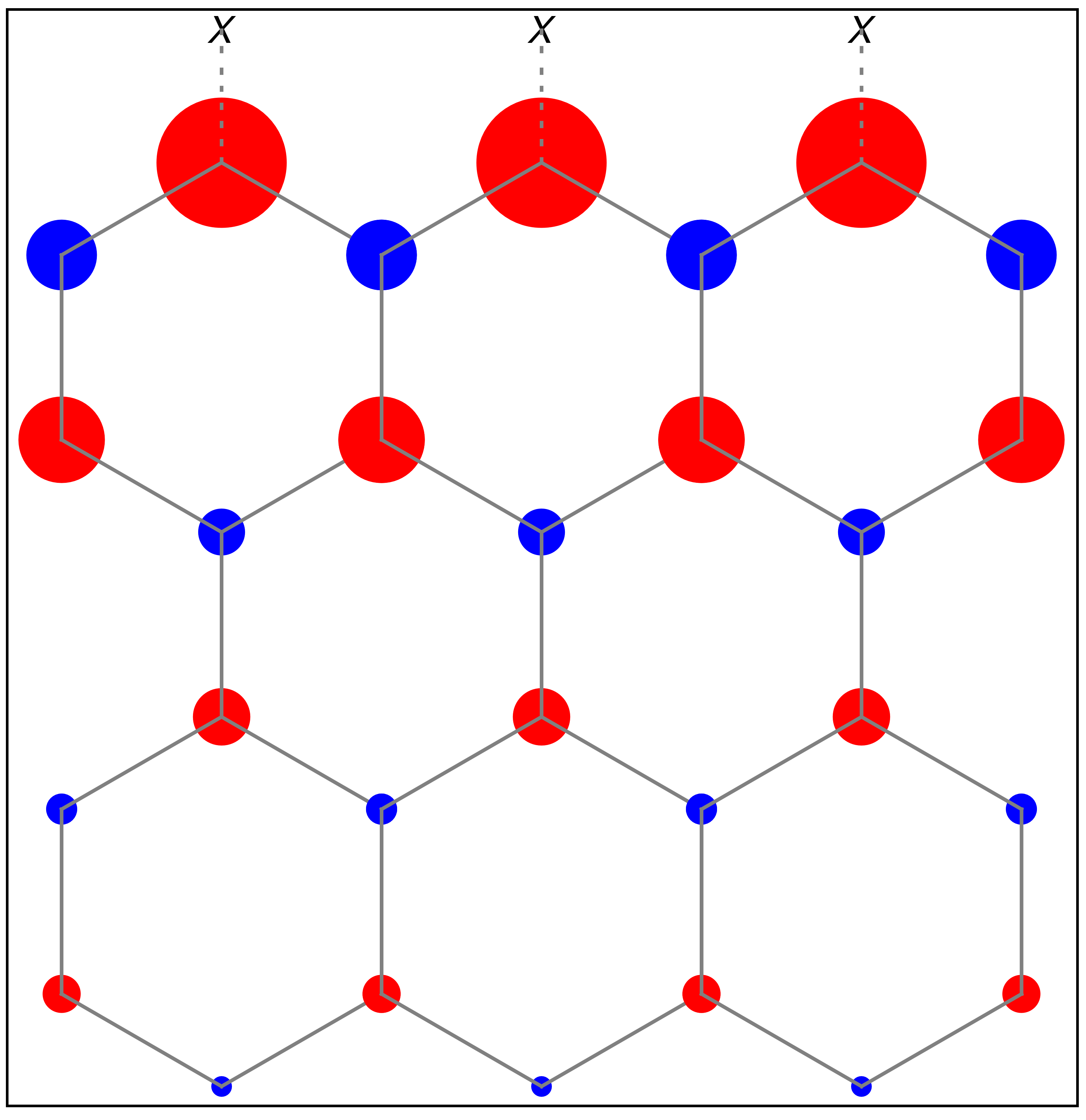}
\par\end{centering}
\caption{{\small{}(Color online) Spin density profile for $k=1.41$ and external
on-site potential a) $\delta_{1}=0.2$ and b) $\delta_{1}=0.4$. For
clarity, the magnitudes on the edge are held constant. The magnitudes
of the spin density are proportional to the radius of each circle
with a phase given by $e^{in\theta_{l}}=\pm1$. \label{fig: Figure3}}}
\end{figure}

\subsubsection{Zig-zag boundary}

\begin{figure}
\begin{centering}
\includegraphics[scale=0.6]{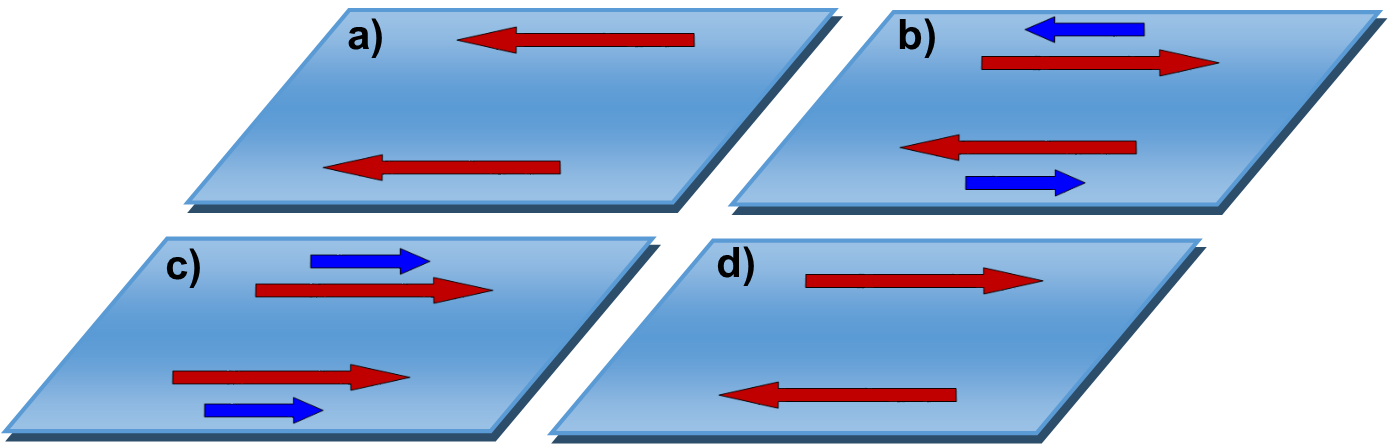}
\par\end{centering}
\caption{{\small{}(Color online) Schematic illustrations of the edge states
propagation with the same momentum $k$. The arrows represent the
edge magnon velocity which depends on the external on-site potential
and the DMI strength. Each figure corresponds to a) single edge magnon
propagating at each boundary in the same direction, b) two edge magnons
with different velocities moving in opposite directions at each boundary,
c) two edge states moving in the same direction at the same boundary
and d) chiral edge states.\label{fig: Figure4}}}
\end{figure}

For a zig-zag boundary, in the absence of external on-site potentials
and zero DMI, the obtained physical solutions are bulk states with
$z^{2}=1$ and $\varepsilon=2\pm\left|J_{1}\right|$. As shown in
the Fig. (\ref{fig:Figure1}a), the outermost $A$ site has two nearest
neighbors and the missing bond generates an attractive potential which
acts as an effective defect which surprisingly does not allow the
formation of edge states. To induce a Tamm-like edge state, the effective
defect is strengthened by turning on the external on-site potential,
$\delta_{1}$. In the Fig. (\ref{fig: Figure2}) the energy spectra
and the decaying factors of the induced edge states are shown for
different values of $\delta_{1}$. As the external on-site potential
is increasing $(\delta_{1}\rightarrow1)$, the branch becomes flatter
{[}Fig. (\ref{fig: Figure2}a){]} and from the edge state wavefunctions,
\begin{equation}
\left(\begin{array}{c}
\psi_{A,n}\\
\psi_{B,n}
\end{array}\right)=z^{n}\left(\begin{array}{c}
z^{-1}\\
\frac{1-\delta_{1}}{J_{2}}
\end{array}\right),\label{eq: Wavefunctions Zigzag D=00003D0}
\end{equation}
 the magnon density is increasingly localized in a single lattice
{[}See Fig. (\ref{fig: Figure3}){]}. In the above equation, the decaying
factor $z$ is a real number. For a wide ribbon\cite{Hatsugai1993,Hatsugai1993a},
the edge state energy spectra is double degenerated and since the
magnon velocity is the slope of the energy spectrum, Fig. (\ref{fig: Figure2}),
the magnons are moving in the same direction at opposite edges, Fig.
(\ref{fig: Figure4}a). Here, as $\delta_{1}$ is increased the slope
is reduced until $\delta_{1}=1$ where the edge state becomes non-dispersive. 

\begin{figure}
\begin{centering}
\includegraphics[scale=0.225]{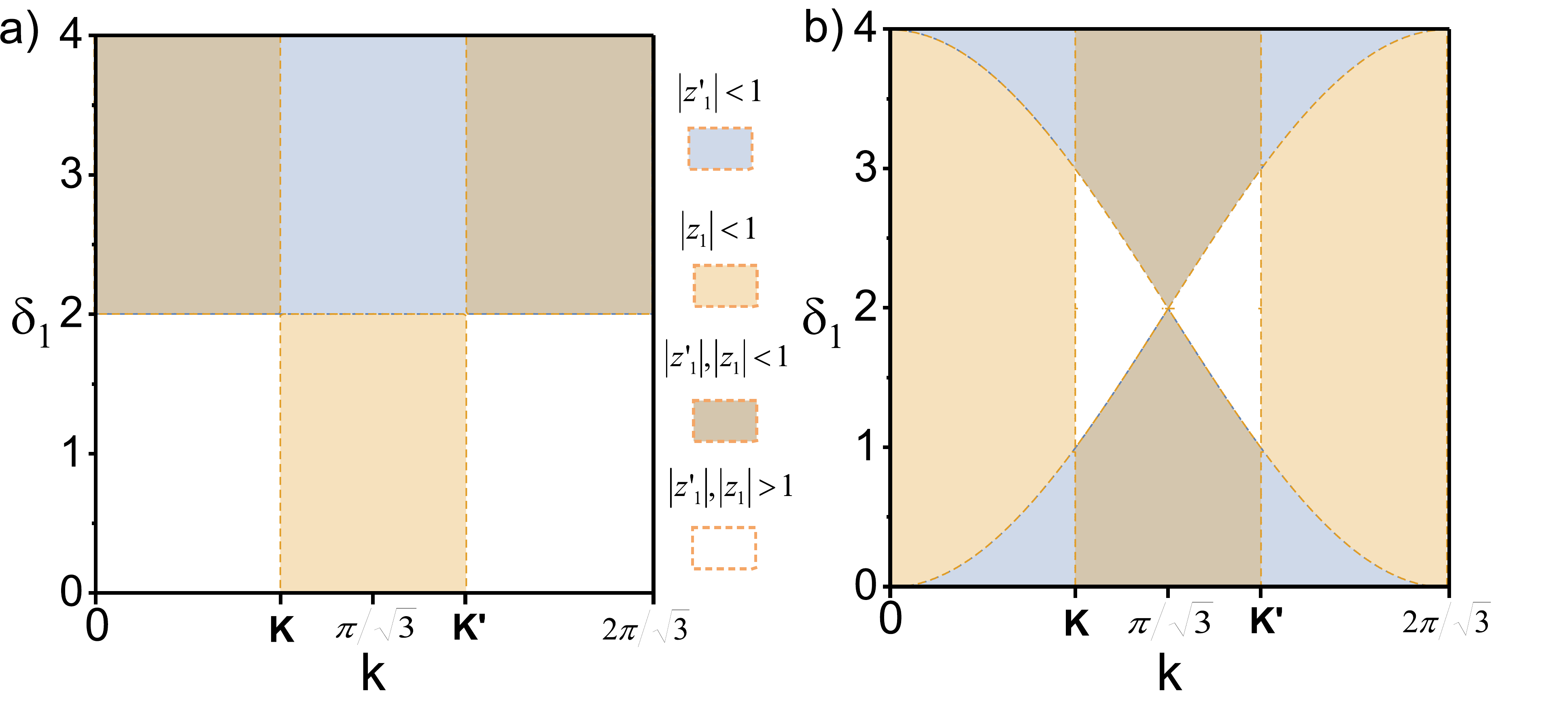}
\par\end{centering}
\caption{{\small{}(Color online) Decaying factor diagram for the edge states
in a honeycomb lattice with a) zig-zag and b) bearded termination
and zero DMI. The number of decaying factors with $\left|z\right|<1$
is the number of edge states in the corresponding region. The (dashed)
lines dividing each region are the points where both decaying factors
reach the unity. \label{fig: Figure5}}}
\end{figure}

If the external on-site potential is increased, in addition to the
shape, the number of edge states can be modified. Depending on the
external on-site potential strength, a zig-zag termination can have
two edge states at each boundary. In the decaying factor diagram,
Fig. (\ref{fig: Figure5}a), each edge state has a corresponding,
$z_{1}$ or $z_{1}^{\prime}$, decaying factor. For $0<\delta_{1}<2$,
there is a single decaying factor between the Dirac points {[}see
also Fig. (\ref{fig: Figure2}){]} and from the Eq. (\ref{eq: Wavefunctions Zigzag D=00003D0})
it is straightforward to show that the edge state in this region is
mainly localized at the $A$ sublattice. For $\delta_{1}>2$, there
are two edge states, the first one, corresponding to $z_{1}^{\prime}$,
is defined over all the Brillouin zone with energy spectra over the
bulk bands (due to the strong external on-site potential). The second
edge state, corresponding to $z_{1}$, is defined in the region $\boldsymbol{K}>k>\boldsymbol{K^{\prime}}$
as in the bearded graphene. Such edge state has a magnon density mainly
localized at the $B$ sublattice with energy spectrum between the
Bulk bands. If the external on-site potential is stronger, $\delta_{1}\gg2$,
the system effectively shows the band structure of a bearded termination
plus a high energy Tamm-like edge state. Moreover, as we mentioned
before, in absence of external on-site potential there are not edge
states. However, for $\delta=2$, there are not edge states either.
This can be observed in the diagram, Fig. (\ref{fig: Figure5}a),
where at such value, $\left|z_{1}\right|=\left|z_{1}^{\prime}\right|=1$
for all values of $k$. At the transition lines (dashed) the modulus
of the decaying factors reaches the unity and the edge states are
indistinguishable from the bulk bands. 

On the other hand, it is well known that the magnon excitations in
a ferromagnetic lattice can be viewed as a synchronic precession of
the spin vectors. Therefore, the sign of the wavefunction, Eq. (\ref{eq: Wavefunctions Zigzag D=00003D0}),
can be related with the spin precession in successive, $n$, rows
and the modulus with the radius of precession which decrease as $n$
increases. If we write the phase of the wavefunction as $e^{in\theta_{l}}=sgn(\psi_{l,n})$,
then, for a given $k$ and $0<\delta_{1}<2$, the synchronic precession
of the spins in successive $n$ rows is in anti-phase ($\theta_{l}=\pi$,
optic-like) if $k<k_{o}$ and in-phase ($\theta_{l}=0$, acoustic-like)
if $k>k_{o}$, here, $k_{0}=\pi/\sqrt{3}$ is the transition point.
Furthermore, at the same $n$, the spins at different sub-lattices
are precessing in anti-phase for $k<k_{o}$ and in-phase for $k>k_{o}$
{[}See Fig. (\ref{fig: Figure3}a){]}. At the transition point $k_{0}$,
the edge state energy is $\varepsilon_{0}=2+\delta_{1}$ and the decaying
factor is zero, Fig. (\ref{fig: Figure2}b). Hence, by the Eq. (\ref{eq: Wavefunctions Zigzag D=00003D0}),
the magnon is completely localized at the edge site, independently
of the external on-site potential strength.

\subsubsection{Bearded boundary}

\begin{figure}
\begin{centering}
\includegraphics[scale=0.22]{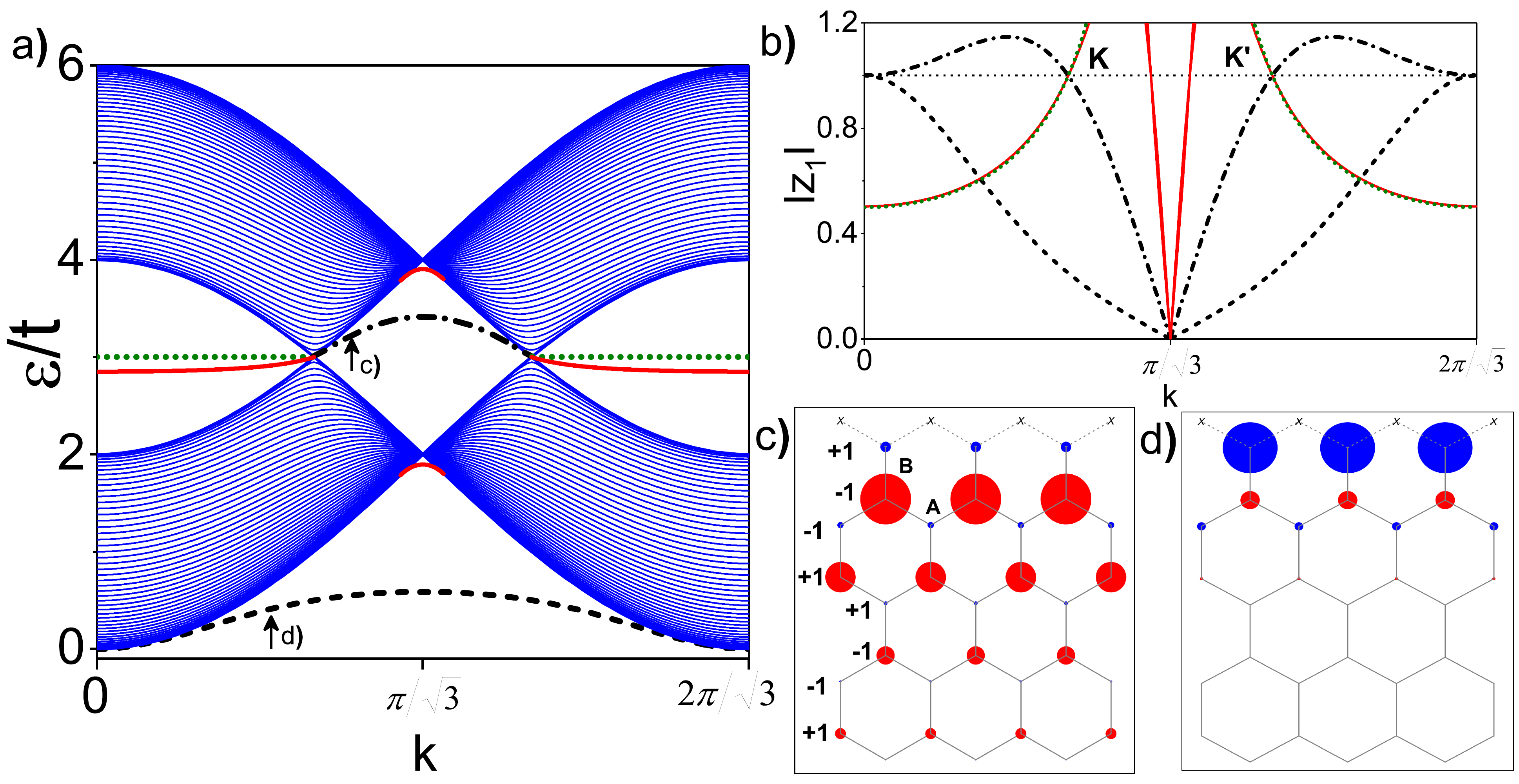}
\par\end{centering}
\caption{{\small{}(Color online) a) Bulk (blue region) and edge state energy
spectra for $\delta_{1}=0$, (black, dashed and dot-dashed lines),
$\delta_{1}=1.8$ (red, continuous lines) and $\delta_{1}=2$ (green,
dotted lines). In b) we show their corresponding decaying factors.
The magnon density profile is shown for the edge states with $\delta_{1}=0$
at c) $k=1.40$ and d) $k=0.96$. Here, the magnitudes of the spin
density are proportional to the radius of each circle with a phase
given by $e^{in\theta_{l}}=\pm1$.\label{fig: Figure6}}}
\end{figure}

We now consider a bearded termination. As shown in the Fig. (\ref{fig:Figure1}b),
the outermost site has two missing bonds and the effective defect
is stronger than the corresponding to a zig-zag boundary. On the contrary
of the fermionic equivalent, the on-site terms provided by the Eq.
(\ref{eq: Bosonic Hamiltonian}) change substantially the edge state
band structure. This is shown in the Fig. (\ref{fig: Figure6}a),
where for $\delta_{1}=0$ there are two edge state energy bands, Eq.
(\ref{eq: Bearded D0 Energy Spectrum}), the first one between the
Dirac points (dot-dashed, black line) and the second one below the
lower bulk bands (dashed, black line). Such edge states are defined
in a region in $k$ completely different to their fermionic equivalent\cite{Klein1994,Jaskolski2011}.
As shown in the Fig. (\ref{fig: Figure6}b), the edge state below
the bulk bands is defined over all the Brillouin zone, except at $k=0,2\pi\sqrt{3}$,
where the decaying factor reaches the unity and the edge state is
indistinguishable from the bulk bands. Excluding those merging points,
the edge state wavefunction with a real decaying factor $z$, 
\begin{equation}
\left(\begin{array}{c}
\psi_{A,n}\\
\psi_{B,n}
\end{array}\right)=z^{n}\left(\begin{array}{c}
\frac{2-\delta_{1}}{J_{1}}\\
z^{-1}
\end{array}\right),\label{eq: Main Bearded D0 wavefunction}
\end{equation}
reveals that the low energy edge state is strongly localized along
the boundary $B$-sites. In the above equation, $z=z_{1}^{\prime}$
, for the edge state below the lower bulk bands and $z=z_{1}$ for
the edge state between the Dirac points {[}see Fig. (\ref{fig: Figure5}){]}.
In the Fig. (\ref{fig: Figure6}c) and Fig. (\ref{fig: Figure6}d),
we plot the magnon density, $\left|\psi_{l,n}\right|^{2}$ for both
edge states at different momentum. Note that the edge states are localized
in different sub-lattices. 

Some interesting features about these edge states are in order here.
From the Fig. (\ref{fig: Figure6}a), for $\delta_{1}=0$ the slope
the edge state energy spectra is positive if $k<k_{0}$ and negative
if $k>k_{0}$. For a wide ribbon, each edge band is doubly degenerated,
hence, the magnons are moving in the same direction (with different
energy) at each boundary, Fig. (\ref{fig: Figure4}c). The fact that
both edge states are strongly localized in different sub-lattices
can be explained if we consider the edge by itself a defect. By a
closer inspection of the wavefunctions, Eq. (\ref{eq: Main Bearded D0 wavefunction}),
the edge state below the lower bulk bands is mainly localized along
the boundary $B$ sites due to the strong attractive potential due
to the missing bonds. The edge state between the bulk bands is mainly
localized along the $A$ sublattice due to the presence of the outermost
$B$ site. In consequence, the outermost $B$ site plays a double
role; acts as an effective defect to host a low energy edge state
and contributes to the formation of the edge state between the Dirac
points. 

The number of edge states is determined by the number of solutions
of the Eq. (\ref{eq: Z solutions for Bearded at D=00003D0}) with
modulus lower than one and the edge state dispersion can be tuned
in all the Brillouin zone with small changes of the external on-site
potential. This can be observed in the decaying factor diagram, Fig.
(\ref{fig: Figure5}b), where the dashed lines separate the regions
in which each edge state is defined. In the region, $0\leq\delta_{1}<1$,
there are always two edge states (for $z_{1}^{\prime}$ and $z_{1}$).
If $\delta_{1}=0$, the first edge state is defined over all the Brillouin
zone, $\left|z_{1}^{\prime}\right|<1$, and the second one between
the Dirac points, $\left|z_{1}\right|<1$. As $\delta_{1}$ is increased
both edge states gradually merge with the bulk bands. For $\delta_{1}=2$,
there is a single edge state with a momentum in the region, $\boldsymbol{K}>k>\boldsymbol{K^{\prime}}$.
This edge state is the flat band in Fig. (\ref{fig: Figure6}a), (dotted,
green line) where the energy spectra closely resembles the fermionic
graphene. If the external on-site potential is increased, $\delta_{1}\gg2$,
the hopping between sites at $n=1$ is almost suppressed and the system
effectively will show the band structure of a zig-zag termination
plus and a high energy Tamm-like edge state along the boundary sites. 

Another important characteristic provided by the explicit form of
the wavefunction, Eq. (\ref{eq: Main Bearded D0 wavefunction}) is
the phase of the spin precession in successive rows. As discussed
in the previous section, the sign of the decaying factor determines
if the phase of the edge state is optic-like or acoustic-like. As
is described in the appendix \ref{sec: Analytical D0}, there are
two decaying factors and their sign reveals that the behavior of the
phase in successive rows is different in both edge states. In particular
for $\delta_{1}=0$, the decaying factor of the edge state connecting
the Dirac points is negative if $k<k_{0}$, then, the spin precession
in successive lattice sites is in anti-phase (optic-like). However,
the decaying factor of the edge state below the lower bulk bands is
positive, if $k<k_{0}$, and the spins in two successive rows are
in-phase (acoustic-like). This provide us two ways to distinguish
these edge states, by their energy and their phase difference in successive
rows. 

Experimentally, the first observation of edge states in a honeycomb
lattice with bearded boundaries were achieved in optical lattices
\cite{Plotnik2014}. Apart from the typical band structure, additional
edge states were observed near the Van Hoove singularities. As is
shown in the Fig. (\ref{fig: Figure6}a) for our model, similar edge
states are obtained for an external on-site potential of $\delta_{1}=1.8$.
Here a nearly flat band plus two highly dispersive edge states near
the Van Hoove singularities (continuous, red lines) are obtained.
As in the reference \cite{Plotnik2014}, the origin of such edge states
is also related to the effective defect generated by the on-site potential
along the boundary sites. In our model, the external on-site potential
is introduced at the outermost sites with fixed hopping terms. However,
as in the case of an square lattice\cite{DeWames1969,Hoffmann1973},
similar physics can be obtained by a renormalization of the hopping
terms along the boundary sites.

\subsection{Non-zero DMI}

A non-zero DMI breaks the lattice inversion symmetry and a non-trivial
gap is induced in the spin-wave excitation spectra. By a topological
approach with the wavefunctions for the infinite system, the Chern
number predicts a pair of counter propagating modes\cite{Owerre2016d}
along the boundary of the finite system. However, the topological
approach does not provide the detailed properties of the edge states
and also does not take into account the on-site potential along the
boundary sites, which, as we will shown in this section, has important
effects in the edge state band structure. 

\subsubsection{Zig-zag boundary}

\begin{figure}
\begin{centering}
\includegraphics[scale=0.226]{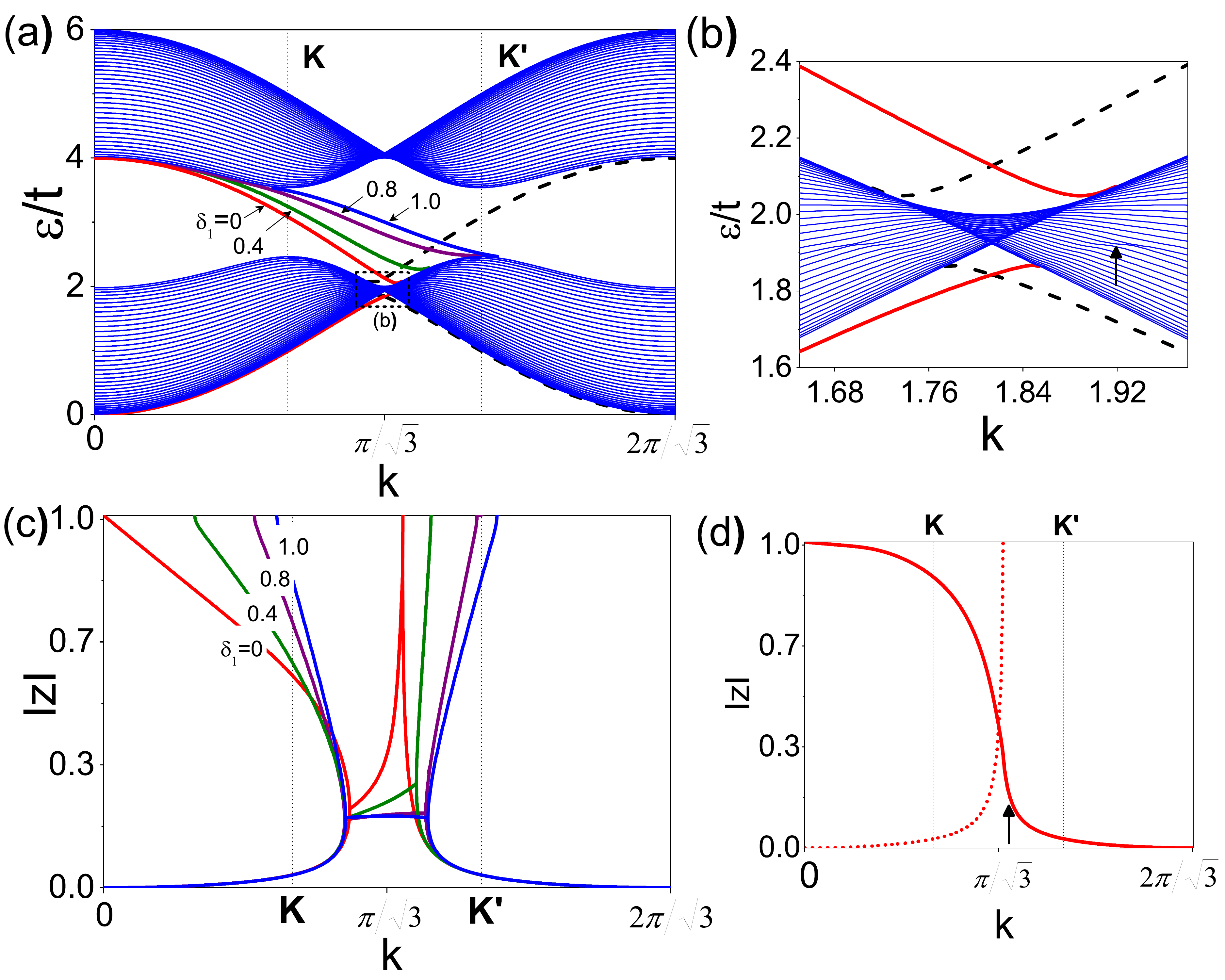}
\par\end{centering}
\caption{{\small{}(Color online) a) Energy spectrum of a zig-zag honeycomb
lattice for $D=0.1\,J$. The lines connecting the upper and lower
bulk bands are the edge states for different values of $\delta_{1}$,
the dashed (black) lines are the edge states at the opposite edge.
As shown in b) for $\delta_{1}=0$ there are additional edge states
below the lower bulk bands, dashed square in a). In c) the decaying
factors of the corresponding edge states in a) is shown. d) Decaying
factors $\left(\delta_{1}=0\right)$ of the edge state below the lower
bulk band in a) and b). The energy spectra in b) reveals an edge state
with energy within the bulk bands (black arrow), the magnitude of
its corresponding decaying factor is given by the arrow in d) \label{fig: Figure7}}}
\end{figure}

We first consider a zig-zag boundary. The energy bands are obtained
by the solutions of the self-consistent Eq. (\ref{eq: Energy autoconsistent zigzag})
with the decaying factors provided by the Eq. (\ref{eq: Quadratic equation general}).
In the Fig. (\ref{fig: Figure7}a) we show the energy bands for a
DMI strength of $D=0.1J$. The blue regions correspond to the bulk
spectra where all the factors, $\left|z_{\nu}\right|=1$. The bands
which transverse the gap are the spectra of the edge states for different
values of $\delta_{1}$. By completeness, we also include the energy
spectra for the edge state at the opposite edge (at large $n$), without
external on-site potential. On the contrary to the predicted by a
topological approach \cite{Owerre2016d,Kim2016a}, the edge state
is not connecting the regions near the Dirac points. As is shown in
the Fig. (\ref{fig: Figure7}a), for $\delta_{1}=0$ (red, continuous
line) the intrinsic on-site potential along the boundary pull the
edge state within the bulk gap to a lower energy region, just over
the lower bulk bands. Furthermore, a new edge state near the Van Hoove
singularities is revealed in the band structure. As is shown in the
zoomed region, Fig. (\ref{fig: Figure7}b), around $k_{0}$ there
are two edge states (at each boundary), over and below the lower bulk
band. The edge state over the bulk bands has a topological origin
and the edge state below is a Tamm-like edge state. 

In general, the edge states depend on two decaying factors, Eq. (\ref{eq: Wavefunctions zig-zag}).
In the Fig. (\ref{fig: Figure7}c) their typical behavior can be observed;
if we move away from $k_{0}$, while one factor decreases to zero
the another one approaches to a critical value (merging point) where
it reaches the unity. In this situation, one component of the edge
state wavefunction becomes an extending wave (bulk wave) and the edge
state is indistinguishable from the bulk bands. However, for the edge
state below the lower bulk bands, (\ref{fig: Figure7}d), in the region
$k>k_{0}$, while one decaying factor reaches the unity the second
one has enough strength to modified the bulk band structure {[}arrows
in Fig. (\ref{fig: Figure7}b) and Fig. (\ref{fig: Figure7}d){]},
in this situation the edge state has energy within the continuum\cite{Zhang2012,Longhi2013}.
For $\delta_{1}=0$ (and $D\neq0$), the edge band within the bulk
gap has a negative slope while the novel edge band below the lower
bulk band has a positive slope. Therefore, the magnons are moving
in opposite directions at the same boundary, Fig. (\ref{fig: Figure4}b).
If the external on-site potential is slightly increased, the Tamm-like
edge magnon merges with the bulk and the magnon propagation will be
like in the Fig. (\ref{fig: Figure4}d). 

As is shown in the Fig. (\ref{fig: Figure7}a) as the external on-site
potential is increasing, the slope of the energy spectra decreases.
In particular, for $\delta_{1}=1$ (uniform case) the energy spectra
closely resembles the fermionic graphene with merging points near
the Dirac points and with the magnons moving in opposite directions
at different boundaries, Fig. (\ref{fig: Figure4}d). In the Fig.
(\ref{fig: Figure7}c) the modulus of the decaying factors is shown
for different values of the external on-site potential. Here, as $\delta_{1}$
is increased, the merging points approaches by the left to the $\boldsymbol{K}$
and $\boldsymbol{K}^{\prime}$ points and the asymmetry around $k_{0}$
is reduced. In the finite region {[}Fig. (\ref{fig: Figure7}c){]}
around $k_{0}$, we have $\left|z_{1}\right|=\left|z_{2}\right|$
and from the Eq. (\ref{eq: Quadratic equation general}) is evident
that the decaying factors are complex conjugates to each other. At
certain momentum both decaying factors become real and they are not
longer identical and, as we mentioned before, while one factor increases
the another one decreases. The region around $k_{0}$, where the edge
states are complex conjugates to each other, is defined for a non-zero
DMI and is located within the bulk gap. Its boundaries in the $k$
space are given by the discriminant of Eq. (\ref{eq: Quadratic equation general})
and is independent of the boundary conditions. If the spectrum of
an edge state crosses this region, their corresponding wavefunction
becomes complex.

\subsubsection{Bearded boundary}

\begin{figure}
\begin{centering}
\includegraphics[scale=0.245]{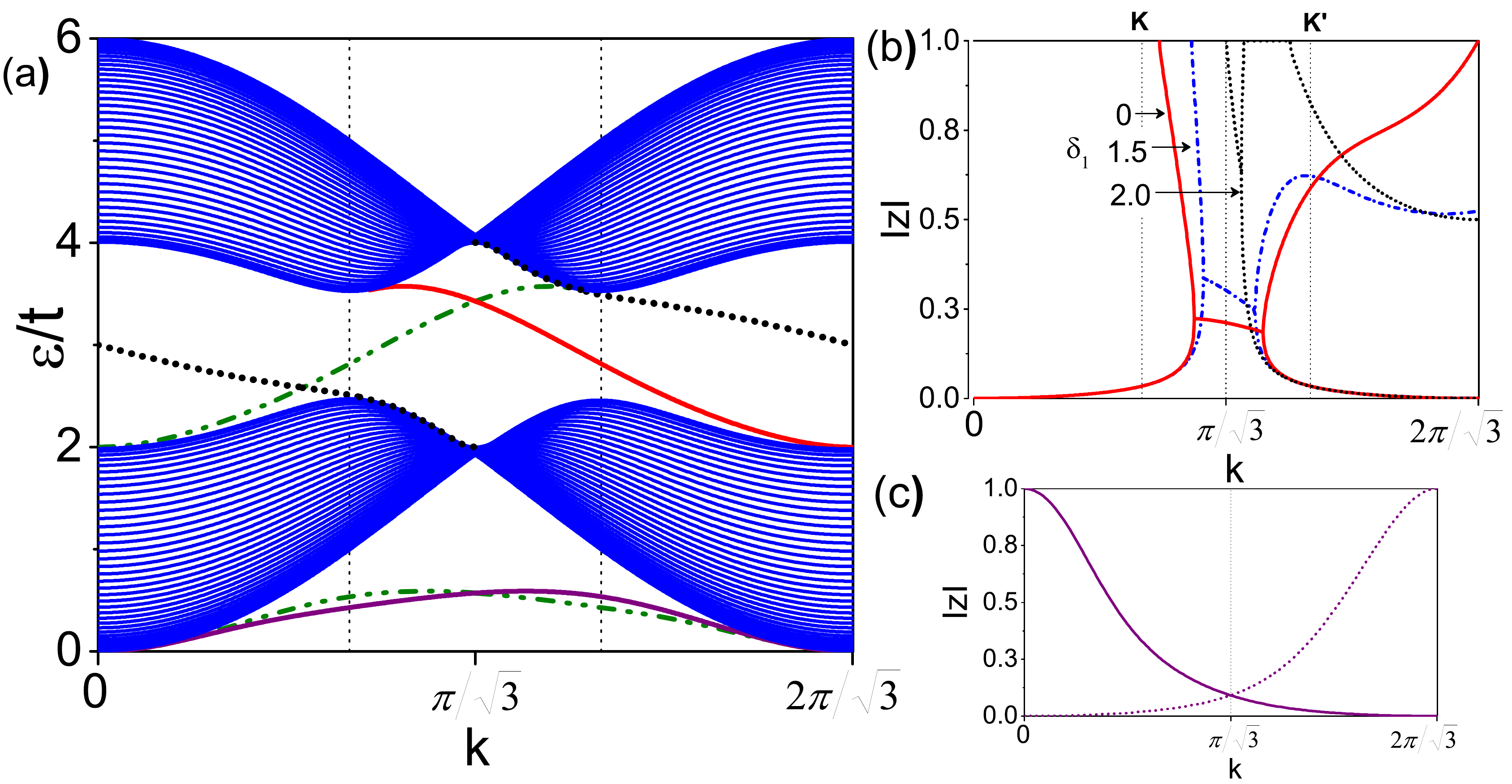}
\par\end{centering}
\caption{{\small{}(Color online) a) Energy spectrum of a bearded honeycomb
lattice. The blue region is the gapped bulk spectra with $D=0.1J$.
For $\delta_{1}=0$, the continuous (red and purple) lines are the
edge states. By completeness, we also include the edge states at the
opposite edge, dot-dot-dashed (green) lines. For $\delta_{1}=2$ there
is a single edge state (black, dotted line). In b) we plot the modulus
of the two decaying factors for the edge state connecting the bulk
bands at different values of $\delta_{1}$. In c) the decaying factors
for $\delta_{1}=0$ are shown for the edge state below the lower bulk
bands. \label{fig: Figure8}}}
\end{figure}

We now consider a bearded termination with a non-zero DMI and arbitrary
external on-site potential. The solutions can be obtained by the self-consistent
equation provided by the Eq. (\ref{eq: General Bearded boundary condition})
and the wavefunctions with the Eq. (\ref{eq: General Wavefunctions Bearded}).
As is shown in the Fig. (\ref{fig: Figure8}) for $\delta_{1}=0$,
there is an edge state crossing the gap (red, continuous line) and
an edge state below the lower bulk bands (purple, continuous line).
Note that the non-zero DMI changes the magnon velocity. In fact, the
edge state within the gap has a negative slope except near the $\boldsymbol{K}$
point where is almost flat. The edge state energy spectrum below the
lower bulk bands has a maximum point where its slope changes. Before
such point and out from the almost flat region, the propagation is
like in the Fig. (\ref{fig: Figure4}b) where, for a fixed momentum
and at the same boundary, the magnons are moving in different directions.
On the other hand, as is shown in the Fig. (\ref{fig: Figure8}a),
to the right of the $\boldsymbol{K}^{\prime}$ point, there are two
edge bands with negative slope (red and purple continuous lines) and
a single edge band with negative slope at the opposite boundary (green,
dot-dot-dashed line), hence the magnons are moving in the same direction
at both edges. 

The effective defect due to the missing bonds is strong in the bearded
boundary, where the edge state energy spectra are distinct to their
fermionic equivalent. As shown in the Fig. (\ref{fig: Figure8}b)
the edge state within the bulk gap (red, continuous line) is defined
in a region to the right of the $\boldsymbol{K}$ point. The edge
state below the lower bulk bands is defined over the whole Brillouin
zone and since its origin is due to the effective defect discussed
in the previous section, it can be shown that is nearly insensitive
to small changes in the DMI strength. In the Fig. (\ref{fig: Figure8}c)
the decaying factors modulus for this edge state are shown. The curves
are almost symmetric around $k_{0}$ and since the decaying factors
are real, the wavefunction decays exponentially to the inner bulk
sites \cite{Doh2013,Pantaleon2017}. 
As discussed in the previous section, as we move away from $k_{0}$,
one decaying factor approaches to the unity while the another one
decreases. Note that the Fig. (\ref{fig: Figure8}b) is similar to
the Fig. (\ref{fig: Figure7}b) except that the plots are tilted to
opposite sides. Here, as the external on-site potential is increased,
the merging points approach to $k_{0}$. In particular for $\delta_{1}=2$,
the edge state has an energy spectrum connecting the Dirac points
(black, dotted line in Fig. (\ref{fig: Figure8}a)). However, in contrast
with their fermionic equivalent, around $k_{0}$ (Van Hoove singularity)
there is an small region in which a high dispersive (and almost indistinguishable)
edge state is also defined, (black-dotted line in Fig. (\ref{fig: Figure8}a)
and (\ref{fig: Figure8}b)). If $\delta_{1}\gg2$, as in the case
for $D=0$, the system effectively will show the band structure of
a zig-zag termination plus and a high energy Tamm-like edge state. 

\section{Concluding remarks \label{sec: Concluding remarks}}

Recently, it has been shown that a system of two-interacting bosons
in a honeycomb lattice satisfy the Hamiltonian given in Eq. (\ref{eq: Bosonic Hamiltonian})
in the limit of strong interaction and with renormalized parameters\cite{Salerno2017}.
In particular, for a bearded boundary, the two-particle system (doublon)
behaves like a single particle and the edge state energy spectra in
the Fig. (\ref{fig: Figure8}) is obtained. In this model, an isotropic
coupling constant $J(>0)$ is considered over the whole semi-infinite
lattice. Due to the on-site contributions in the Hamiltonian, Eq.
(\ref{eq: Bosonic Hamiltonian}), the missing bonds reduce of the
on-site energy along the boundary sites. Since the on-site contribution
is the number of nearest-neighbors, the external on-site potential
$\delta_{1}$ is introduced as an \textit{extra bond} which does not
have any effect in the hopping parameters. The model presented in
this paper has the advantage that the boundary conditions can be easily
modified without modify the Harper's equation. For example, instead
of introduce an extra bond, a renormalization of the hopping parameters
along the boundary can also be introduced by allowing to the exchange
parameters to deviate from the bulk values. Similarly to the model
for the surface spin waves in an square lattice\cite{DeWames1969},
we can consider different coupling constants along the boundary sites.
However, such modifications does not changes the physics of this problem
where the edge by itself is considered as an effective defect. As
we mentioned before, such effective defect is due to the missing bonds
along the boundary sites and is a characteristic of the bosonic lattices,
consequently, the physical systems modeled by a bosonic Hamiltonian,
Eq. (\ref{eq: Bosonic Hamiltonian}), may show edge states naturally
or through the application of small external on-site potentials. 

The interesting properties of the honeycomb lattice may be experimentally
accessible through engineered spin structures on metallic surfaces\cite{Banerjee2016a},
using ultra-cold bosonic atoms trapped in honeycomb optical lattice
\cite{Vasic2015}, photonic lattices\cite{Polini2013,Bellec2014}
and so forth. Therefore, the distribution of the edge magnons, the
spin-density and their dependence with the DMI strength and external
on-site potentials presented in this paper could be useful for experiments
in small sized mono-layers, thin film magnets or artificial lattices.

\section{Conclusions\label{sec: Conclusions}}

We have studied the on-site potential effects in the magnon edge states
in a honeycomb ferromagnetic lattice with zig-zag and bearded boundaries.
For zero DMI, the connection of the formation of the Tamm-like edge
states with the effective defect due to the on-site potential along
the outermost sites has been elucidated. For non-zero DMI, we found
that the edge state energy spectra is modified due to the missing
bonds along the boundary sites and their distribution in the momentum
space is different to that predicted by a topological approach. For
both zig-zag and bearded boundaries and for zero and non-zero DMI,
the edge state properties are discussed and Tamm-like edge states
have been revealed. Nevertheless, if these edge states can also be
predicted by a topological approach is still an open question. We
found that the Tamm-like and the topologically protected edge states
are tunable by modifying the external on-site potential and the DMI.
Furthermore, analytical expressions for the edge state energy spectrum
and their corresponding wavefunctions are obtained. We believe that
our results may explain the unconventional edge states recently found
in optical \cite{Plotnik2014,Mili2017} and acoustic \cite{Ni2017}
lattices and motivate new experiments in topological bosonic insulators. 

Note added.\textSFx{} After completion of this work, a related work
have appeared in which the magnon edge states in a honeycomb ferromagnet
are also discussed \cite{Pershoguba2017}. The analytical solutions
and the edge state properties, however, are not discussed. 

\appendix

\section{Analytical solutions for $D=0$\label{sec: Analytical D0}}

In this appendix, we derive the edge state energy spectrum and wavefunctions
for a semi-infinite ferromagnetic honeycomb lattice with a bearded
boundary, in absence of DMI and with an arbitrary external on-site
potential $\delta_{1}$. From the Eq. (\ref{eq: Quadratic equation general})
for $D=0$, the characteristic equation of $H_{ef}$, Eq. (\ref{eq: Effective Hamiltonian for the edge state}
), is given by,
\begin{equation}
\left(3-\varepsilon\right)^{2}-J_{1}^{2}-J_{2}^{2}-J_{1}J_{2}\left(z+z^{-1}\right)=0.\label{eq: Characteristic equation D=00003D0}
\end{equation}
For a fixed $k$, the above equation relates the decaying factor $z_{1}$
with the energy $\varepsilon$. However, an additional equation provided
by the boundary conditions is required. From the Harper's equation,
Eq. (\ref{eq:  Harper Equation General}), with the replacements $J_{1}\rightarrow J_{2}$
, $J_{2}\rightarrow J_{1}$ and taking into account the missing bonds,
the additional equation for the edge state at $n=1$ is written as,
\begin{equation}
\left(3-\varepsilon\right)\left(1+\delta_{1}-\varepsilon\right)-J_{1}J_{2}z+J_{2}^{2}=0.\label{eq: Additional Bearded D0}
\end{equation}
Here, both Eq. (\ref{eq: Characteristic equation D=00003D0}) and
Eq. (\ref{eq: Additional Bearded D0}) provide us a complete set of
equations for the decaying factor and the energy spectrum. Therefore,
for an arbitrary external on-site potential, $\delta_{1}$, the decaying
factor satisfy, 
\begin{equation}
az_{1}^{2}+bz_{1}+c=0,\label{eq: Z solutions for Bearded at D=00003D0}
\end{equation}
where, $a=\left(-2+\delta_{1}\right)^{2}J_{2}$, $b=J_{1}\left[\left(-2+\delta_{1}\right)^{2}-J_{1}^{2}\right]$
and $c=-J_{1}^{2}J_{2}$. Explicitly, 
\begin{equation}
z_{1}=\frac{-\left(\delta_{b}^{2}-J_{1}^{2}\right)^{2}J_{1}\pm\left|J_{1}\right|\sqrt{\left(\delta_{b}^{2}-J_{1}^{2}\right)^{2}+4\delta_{b}^{2}}}{2\delta_{b}^{2}},\label{eq: Bearded D0 Z solution}
\end{equation}
where $\delta_{b}=-2+\delta_{1}$ and $J_{2}=1$. On the other hand,
the edge state energy spectrum satisfy, 
\begin{equation}
a_{1}\varepsilon_{r}^{2}+b_{1}\varepsilon_{r}+c_{1}=0,\label{eq: E solutions for Bearded at D=00003D0}
\end{equation}
where, $\varepsilon_{r}=(\varepsilon-3)-(-2+\delta_{1})$, $a_{1}=(-2+\delta_{1})J_{1}$,
$b_{1}=b$ and $c_{1}=-(-2+\delta_{1})J_{1}J_{2}^{2}$. The edge state
energy spectra have two solutions given by, 
\begin{equation}
\varepsilon=\frac{6\delta_{b}+\delta_{b}^{2}+J_{1}^{2}\pm sgn\left(J_{1}\right)\sqrt{\left(\delta_{b}^{2}-J_{1}^{2}\right)^{2}+4\delta_{b}^{2}}}{2\delta_{b}}.\label{eq: Bearded D0 Energy Spectrum}
\end{equation}
From the above equation and by a closer inspection of the decaying
factors, Eq. (\ref{eq: Bearded D0 Z solution}) two edge states can
be defined. The wavefunction satisfying the boundary condition, 
\begin{equation}
\left(2-\delta_{1}\right)\psi_{B,1}-J_{1}\psi_{A,0}=0,\label{eq: Bearded edge condition D=00003D0}
\end{equation}
can be written as, 
\begin{equation}
\psi_{l,n}=z_{1}^{n}\left(\begin{array}{c}
\frac{2-\delta_{1}}{J_{1}}\\
z_{1}^{-1}
\end{array}\right),\label{eq: Bearded D0 wavefunction}
\end{equation}
where the decaying factor $z_{1}$ is given the by Eq. (\ref{eq: Bearded D0 Z solution}).
At the point $k_{0}=\pi/\sqrt{3}$, the edge states are completely
localized at the boundary sites with energy,
\begin{equation}
\varepsilon_{k_{0}}=\frac{1}{2}\left(6+\delta_{b}\right)\pm\sqrt{4+\delta_{b}}.\label{eq: Energy D0 at kpi3}
\end{equation}
In particular, for $\delta_{1}=2$ the Eq. (\ref{eq: Z solutions for Bearded at D=00003D0})
provide us a single decaying factor, $z_{1}=-J_{2}/J_{1}$ and the
Eq. (\ref{eq: E solutions for Bearded at D=00003D0}) a single solution
$\varepsilon=3$, which is a flat band similar to the fermionic graphene.
Following the same procedure, the analytical form of the decaying
factor and the edge state energy spectrum for a zig-zag boundary can
also be obtained. 

\bibliographystyle{apsrev4-1}

\end{document}